# DPA on Quasi Delay Insensitive Asynchronous Circuits: Formalization and Improvement

G.F. Bouesse, M. Renaudin, S. Dumont

TIMA Laboratory, CIS Group
26 Av. Félix Viallet,
38031 Grenoble Cedex
e-mail: fraidy.bouesse@imag.fr

Fabien Germain

SGDN/DCSSI
51 bd. De la Tour Maubourg
75700 Paris Cedex07
e-mail: fabien.germain@sgdn.pm.gouv.fr

**Abstract**

*The purpose of this paper is to formally specify a flow devoted to the design of Differential Power Analysis (DPA) resistant QDI asynchronous circuits. The paper first proposes a formal modeling of the electrical signature of QDI asynchronous circuits. The DPA is then applied to the formal model in order to identify the source of leakage of this type of circuits. Finally, a complete design flow is specified to minimize the information leakage. The relevancy and efficiency of the approach is demonstrated using the design of an AES crypto-processor.*

## I. Introduction and motivations

Since the discovery of the Power Analysis attacks such as Single Power Analysis (SPA) and Differential Power Analysis (DPA), asynchronous logic has been presented as a new alternative design solution against side channels attacks.

In fact, Power Analysis attacks with DPA firstly introduced in 1998 by Paul Kocher [1] use the weakness of chip hardware implementations and software running on cryptographic devices (particularly smart card), to reveal the chip's confidential information. Secret keys are removed from device by observing and monitoring the electrical activity of a device and performing advanced statistical methods.

Additionally to their absence of clock signal which demonstrates the practical way to eliminate a global synchronization signal, asynchronous logic is well-known for its ability to decreasing the consumption and to shape circuits' current. [2][3][4] demonstrate how 1-of-n encoded speed-independent circuits improve security by eliminating data dependent power consumption. Symmetry in data communication and data processing, persistent storage, timing information leakage and propagating alarm signals (to defend chip against fault induction) are design aspects addressed by those papers for increasing chip resistance. The countermeasures that used the Self-timed circuit properties are all focused on balancing the operation through special DI Coding scheme. The concrete results of these approaches which illustrates how is Quasi Delay Insensitive asynchronous logic factor effective for resisting against DPA has been presented in [5]. By investigating and analyzing three different QDI DES architectures and design styles, it is demonstrated how the properties of 1-of-N encoded data and four-phase handshake protocol significantly improve the DPA resistance. However this study has also pointed out the limits of this design approach methodology by showing some residual sources of leakage.

This paper is focused on these residual sources of leakage that are still observable when implementing a DPA attack on secured QDI asynchronous circuits as described in [5]. The objective of this paper is to formalize a model of the current dissipated in this type of circuit in order to identify the sources of these remaining leakages and thus propose a new design flow to reduce them.

The paper is organized as follows. Section II presents the asynchronous properties that are used to design a DPA resistant chip, especially N-rail quasi delay insensitive asynchronous logic together with four-phase protocol. Section III investigates the electrical current model of this type of circuit and then section IV recalls the DPA attack skill by applying the attack on the model. Validations are presented in section V. The improvement of the design flow and results are reported in Section VI. The paper is concluded by giving some design perspectives in section VII.

## II. Asynchronous logic and DPA

Asynchronous circuits represent a class of circuits which are not controlled by a global clock but by the data themselves. In fact, an asynchronous circuit is composed of individual modules which communicate to each other by means of point-to-point communication channels. Therefore, a given module becomes active when it senses the presence of incoming data. It then computes them and sends the result to the output channels. Communications through channels are governed by a protocol which requires a bi-directional signalling between senders and receivers (request and acknowledge). They are called Handshaking protocols [6] which are the basis of the sequencing rules of asynchronous circuits.

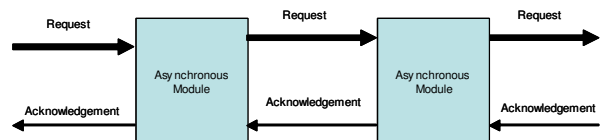

*Fig. 1: Handshake based communication between modules. A module can actually be of any complexity.*

There are two main classes of handshaking protocols: two-phase protocol and four-phase protocols. In this work, only the four-phase protocol is considered and described.

- Four-phase protocol

This protocol requires a return to zero phase for both data/requests and acknowledgements.

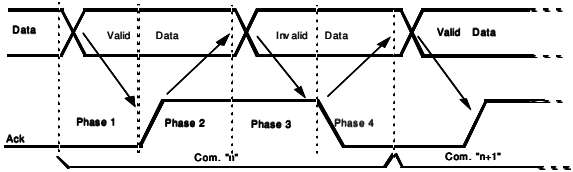

*Fig. 2: Four-phase handshaking protocol*

Phase 1: Data detection (invalid Data to valid Data)
Phase 2: Acknowledgement is set to one
Phase 3: Data are re-initialized (valid Data to invalid Data, return to zero phase)
Phase 4: Acknowledgement is reset (return to zero phase)

Contrary to synchronous circuits where the shape of the current (current peaks) depends on the previous states and data values, QDI asynchronous logic using a four-phase protocol re-initializes all previously activated nodes before processing a new data. This behaviour enables the designer to precisely control the transitions involved in a given computation (fig. 3). Moreover, because it is based on hazard free logic QDI asynchronous circuits eliminate all current variations caused by glitches.

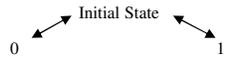

*Fig. 3: Types of transitions*

- Signalling

As presented above, the implementation of a four-phase handshaking protocol requires sensing the presence of data in phase 1 and phase 3. In order to do so, dedicated logic and special encoding are necessary for sensing data validity/invalidity and for generating the acknowledgement signal.

- Data/ Request encoding

Considering that one bit has to be transferred through a channel using the four phase protocol, one bit has to encode three different values: invalid, valid at '1', valid at '0'. Two wires (A0, A1) are then required to encode the three states. This technique is called dual-rail encoding (table1).

- Acknowledge / Completion signal generation

The acknowledgement signal is generated by taking advantage of the data-encoding. As depicted in fig. 4, a Nor gate is usually used to sense the dual-rail encoding output for generating the completion signal.

*Table 1: Dual rail encoding of the three states required to communicate 1 bit.*

| Channel data | A0 | A1 |
|---|---|---|
| 0 | 1 | 0 |
| 1 | 0 | 1 |
| Invalid | 0 | 0 |
| Unused | 1 | 1 |

Dual rail encoding is easily extended to N rails. It is called 1-of-N encoding. This encoding data scheme is useful to reduce the number of electrical transitions involved in a given computation which reduces the power consumption. For the sake of DPA resistance, 1-of-N encoding ensures that the same number of transitions is required to encode the values 0 to N-1.

- Balanced data path

As an example, consider the xor function. Fig. 4 shows a dual-rail xor gate implementation. Every computations of this dual-rail xor gate involve a fixed and constant number of transitions regardless of the data values. Hence, the opportunity to have data independent power consumption i.e. not correlated to the processed data is exactly the goal to achieve for DPA resistant chip.

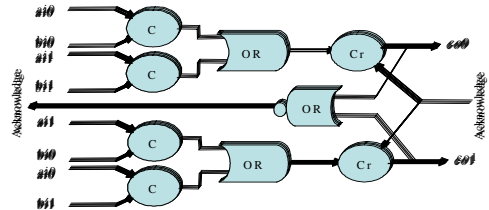

*Fig. 4: Dual-rail xor gate with four-phase handshake protocol. Dual rail "co" outputs the xor function performed between dual rail inputs "ai" and "bi".*
*(Cr is a Muller gate with a reset signal)*

The Muller gate (C-element) generates an up-transition when up-transitions occur at all its inputs, and generates a down-transition when down-transitions occur at all its inputs. The Muller C-element's truth table and symbol are given in Fig. 5.

$$Z = XY + Z(X+Y)$$

| X | Y | Z |
|---|---|---|
| 0 | 0 | 0 |
| 0 | 1 | Z-1 |
| 1 | 0 | Z-1 |
| 1 | 1 | 1 |

*Fig. 5: Muller gate or (C-element)*

However, the QDI implementation of a function is not always balanced. In such cases, the gate structure is modified to ensure that all data paths and control paths are balanced and do involve a constant number of transitions [7].

Publication [5] demonstrated that secured QDI asynchronous logic permit to achieve a very high level of DPA resistance. However, it also showed that place and route steps are unbalancing some paths, thus creating sources of leakage. Let's now precisely analyze this point using a formal approach.

## III. Electrical model of secured QDI asynchronous logic

In this section, we propose a current model of a QDI block implementing all the secure techniques described in section II. We assume that all blocks of the circuit are balanced so that a fix number of logical transitions are required regardless of the input data.

Let us first estimate the power dissipation of a logical block. Power dissipation in static CMOS gate has the following components:

1. Static power dissipation due to junction leakage current, $P_l$.
2. Short-circuit power dissipation, $P_s$.
3. Dynamic power dissipation, $P_d$.

As Static and Short-circuit power represent about 15% of the CMOS gate power dissipation, we consider for the purpose of this study the Dynamic power dissipation ($P_d$) which is defined as the power required to charge and discharge the capacitive load of the gate. Its expression is given by:

$$P_d = \eta f C V_{dd}^2 \qquad (1)$$

Where $\eta$ is a switching activity ratio, $f$ is the switching frequency, $V_{dd}$ is the supply voltage, and $C$ is the total charge of the output gate node, defined by: $C=C_l+C_{par}+C_{sc}$ in which $C_l$, represents the load capacitance (gate and routing capacitance), $C_{par}$ is the parasitic capacitance, and $C_{sc}$ is the Short-circuit equivalent capacitance. This model of power dissipation can be extended to a CMOS gate working in a QDI asynchronous environment.

$$P_{da} = \eta f_a C V_{dd}^2 \qquad (2)$$

Where $f_a$ is the switching frequency of the acknowledge signal. The fix number of logical transitions in each QDI asynchronous block offers the possibility to estimate the block's dynamic power dissipation. Let $N_t$ be a number of logical transitions ($N_t$ is fixed for each block). The block's dynamic power dissipation is given by:

$$P_{db} = \eta f_a \sum_{i=1}^{N_t} C_i V_{dd}^2 \qquad (3)$$

The aim of the asynchronous balanced paths is to always have the same current profile regardless of the data computed. This modeling of the power dissipation shows that in spite of having the same number of transitions, the loading charge of each gate is not necessarily equal. Let then estimate the dynamic current dissipated in a block. The gate dynamic power dissipation is given by:

$$I(t) = C \frac{dV}{dt} \qquad (4)$$

Let define $N_c$ as the number of gates along the critical data-path. $N_c$ represents the maximum number of gate in series in the block. This number ($N_c$) can be used to divide a block in $N_c$ logical levels. $N_{ij}$ is the number of gate switching at each logical level ($N_c$). The expression of the block dynamic current profile is now expressed by:

$$P_{dc}(t) = \sum_{i=1}^{N_c} \left[ \sum_{j=1}^{N_{ij}} I_{ij}(t) \right] + P_{dn}(t) \qquad (5)$$

$I_{ij}(t)$ represents the dynamic current of the $jth$ gate of level $i$ and $P_{dn}$ is a dynamic noise function. The values $N_t$, $N_c$ and $N_{ij}$ are determined by a simple analysis of a graphic representation of the block. For example, let consider the block of fig. 4. Its representation in the form of a directed graph is presented in fig. 5 ($G_{xor}=(V,E)$). The directed graph $G_{xor}(V,E)$ is built from the gate Netlist by defining all the gates as the elements of the set V (vertices) and all the interconnections as the elements of the set E(directed edges).

The graphic representation adopted in this study is well appropriated for this type of analysis. In the one hand, it offers the opportunity to formally verify the logical symmetry of the data-path and in the other hand it offers the possibility to annotate the graph with information collected at each different phase of the design.

$M_i$, $O_i$, $H_i$ and $N_i$ are annotated with all gates' parameters and edges $E_i$ are annotated with all nets' parameters.

These annotations after the back end step permit to take into account logical and real physical elements in the graph analysis.

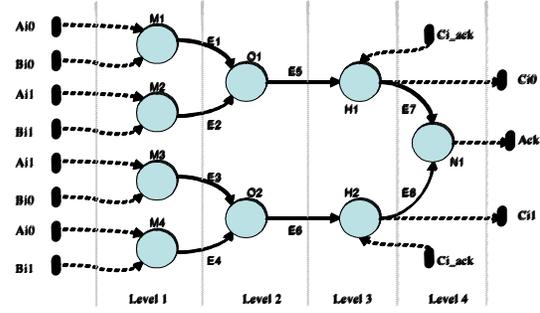

*Fig. 5: Annotated directed graph $G_{xor}(V,E)$ of the Dual-Rail gate Xor.*

All dotted lines represent inputs and outputs of the block. We deduce through the graph exploration the values of $N_t$, $N_c$ and $N_{ij}$:

$$N_t = N_c = 4 \; ; \; N_{1j}=N_{2j}=N_{3j}=N_{4j}=1$$

Therefore,

$$P_{dcxor}(t) = (I_{11}(t)+I_{21}(t)+I_{31}(t)+I_{41}(t)) + P_{dn}(t) \qquad (6)$$

Equation (6) represents, in a first approximation, the profile of the dynamic current of the Dual-rail xor gate. This approach can be extended to all secured QDI asynchronous block.

This formal model enables to evaluate the sensitivity of the secured QDI asynchronous circuits to DPA with high accuracy.

## IV. Applying DPA on the formal model

Before applying DPA attack on the formal model, we first review the basis of the attack. The formalization adopted in this paragraph was introduced by Thomas S. Messerges and Al. in [7].

DPA attack is performed by computing $N$ random values of plain-text-input ($PTI_i$). For each of the $N$ plain-text-input, a discrete time power signal $S_{ij}$ and cipher-text-output are collected. The index $i$ of power signal $S_{ij}$ corresponds to the $PTI_i$ that produced the signal and the $j$ index corresponds to the time of the sample. According to a DPA algorithm, the $S_{ij}$ are split into two sets by a separating function $D$.

$$S_0 = \{S_{ij}|D=0\} \qquad S_0 = \{S_{ij}|D=1\} \qquad (7)$$

The average power signal of each set is given by:

$$A_0[j] = \frac{1}{|n_0|}\sum_{i=1}^{n_0} S_{ij} \qquad A_1[j] = \frac{1}{|n_1|}\sum_{i=1}^{n_1} S_{ij} \qquad (8)$$

Where $|n_o|$ and $|n_1|$ represent the number of power signals $S_{ij}$ respectively in set $S_0$ and $S_1$. The DPA bias signal is obtained by:

$$T[j] = A_0[j] - A_1[j] \qquad (9)$$

If the DPA bias signal shows important peaks, it means there is a strong correlation between the $D$ function and the power signal. Selecting an appropriate *D* function is then essential in order to guess a good secret key. The methods to succeed the attack with a minimum of random values are presented in [8]. An example of such a *D* function is as follows:

1- DES algorithm:
$$D(C_1, P_6, K_0) = SBOX1(P_6 \oplus K_0)(C_1)$$
Where $C_1$ = first output bit of *SBOX1* function.
$P_6$ = 6-bit plain-text-input of the *SBOX1* function.
$K_0$ = 6-bit of the first round of *SBOX1*.
*SBOX1* = a substitution function of DES with 4-bit output.

2- AES algorithm:
$$D(C_1, P_8, K_8) = XOR(P_8, K_8)(C_1)$$
Where $C_1$ = first output bit of *XOR* function.
$P_8$ = 8-bit plain-text-input of the *XOR* function.
$K_8$ = 8-bit of the first round of *XOR*.
*XOR* = a xor function of AES with 8-bit output.

The number of bits chosen for $C_i$ in the selection function determinates the number of sets to create. If only one bit is chosen (which is the case mostly used), two sets are created as show in equation (7).

Let us apply this technique to a secured QDI asynchronous design. Choosing a XOR for the *D* function implies to analyse the electrical signature of an XOR gate. This function is chosen because it directly handles the secret key in most of the cryptographic algorithms.

Contrary to synchronous design where the DPA attack reveals path dissymmetry of the attacked bit ($C_i$), DPA on the secured QDI asynchronous design reveals path dissymmetry of all rails that are used to encode the attacked bit. In fact, applying DPA on Dual-rail xor gate requires comparing the electrical behaviour of paths which compute rail $C_{O0}$ and rail $C_{O1}$. Then, the average current signal of both sets of equation (8) is written as follows:

$$A_{xor0}[t] = \frac{1}{2}(I_{11}(t) + I_{12}(t) + I_{21}(t) + I_{31}(t) + I_{41}(t) + I_n(t))$$
$$A_{xor1}[t] = \frac{1}{2}(I_{13}(t) + I_{14}(t) + I_{22}(t) + I_{32}(t) + I_{41}(t) + I_n(t))$$
(10)

Where $I_n(t)$ is a noise signal. The electrical signature is given by:

$$S[t] = T[t] = \left(C_{11}\frac{dVout_{11}}{dt_{11}} + C_{12}\frac{dVout_{12}}{dt_{12}} + C_{21}\frac{dVout_{21}}{dt_{21}} + C_{31}\frac{dVout_{31}}{dt_{31}} + C_{41}\frac{dVout_{41}}{dt_{41}}\right)$$
$$- \left(C_{13}\frac{dVout_{13}}{dt_{13}} + C_{14}\frac{dVout_{14}}{dt_{14}} + C_{22}\frac{dVout_{22}}{dt_{22}} + C_{32}\frac{dVout_{32}}{dt_{32}} + C_{41}\frac{dVout_{41}}{dt_{41}}\right)$$
(11)

as $\frac{dVout_{ij}}{dt_{ij}} \cong \frac{\Delta V}{\Delta t_{ij}}$ this expression becomes:

$$S[t] = \Delta V \left(\frac{C_{11}}{\Delta t_{11}} + \frac{C_{12}}{\Delta t_{12}} - \frac{C_{13}}{\Delta t_{13}} - \frac{C_{14}}{\Delta t_{14}}\right) + \Delta V \left(\frac{C_{21}}{\Delta t_{21}} - \frac{C_{22}}{\Delta t_{22}}\right) + \Delta V \left(\frac{C_{31}}{\Delta t_{31}} - \frac{C_{32}}{\Delta t_{32}}\right)$$
(12)

$\Delta t$ represents the physical time taken by the gate to charge/discharge its output node. This time depends on the value of *C*. Recalling that $C = C_l + C_{par} + C_{sc}$.

Equation (12) clearly demonstrates that regardless of the symmetry of the logical data-path (same number of transitions in each data-path), the differential current analysis of two symmetric data-paths reveals the effects of each gate's charge capacitance (*C*). It formally illustrates the impact of the gate's charge, particularly the load capacitance, on the DPA bias signal.

## V. Validation using electrical simulations

All electrical simulations are performed with Eldo and used the HCMOS9 design kit (0.13µm) from STmicroelectronic. The electrical simulation offers the possibility to analyze without disturbing signal (noise), the gate's electrical behaviour with more details. Hence, the number of necessary messages (*N*) is minimal.

Figure 6 illustrates the electrical signature of the dual-rail xor gate using all load capacitances $C_{lij}$ equal ($C_{lij}$ represents the load capacitance of the *jth* gate of level *i*). Both evaluation and return to zero phases are analyzed. Signal *S(t)* shows a few peaks due to internal gate capacitance: Short-circuit capacitance ($C_{sc}$) and parasitic capacitance ($C_{par}$).

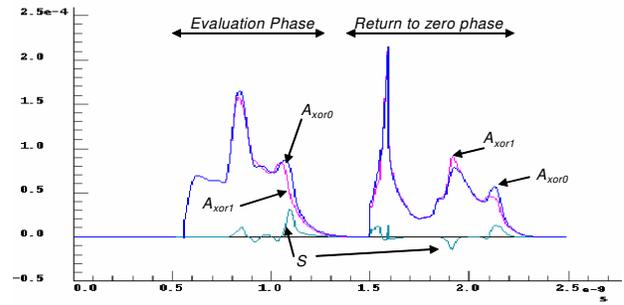

*Fig. 6: Electrical signature of a Dual-rail xor gate.*
$C_{lij}$ = 8 femto farad.

Let us then vary the value of the load capacitance (precisely the value of the interconnection capacitance) in order to evaluate its effect on the gate electrical signature (Fig. 7). A default ($C_d$) value of net capacitance is fixed to 8fF (femto Farad).

Fig. 7-a represents the profile of signal *S(t)* when $C_{l31}$ is two times bigger than $C_d$. As $C_{l31}$ is on the third level, we have one important peak at the end of each phase. It corresponds to the last term of the equation (12). When $C_{l21}$ is fixed to $2*C_d$, two important peaks appear in the signal *S(t)* (fig. 7-b). In fact, as this value is inside the data-path, all computing operations after this gate are shifted by the time taken to charge/discharge this node. It is confirmed when $C_{l11}$ and $C_{l12}$ are both two times bigger than $C_d$ (fig. 7-c). This dissymmetry is amplified when the difference of capacitance between both data-paths increases. In fig. 7-d the values of $C_{l12}$ and $C_{l11}$ are four times bigger than $C_d$. As the difference occurs in the beginning of the block, the electrical curve of both sets are completely shifted so that the electrical signature is maximum.

This analysis confirmed the strong dependence between the net capacitance and the DPA bias signal of the balanced QDI asynchronous design. This directly points out the necessity to precisely control the place and route step.

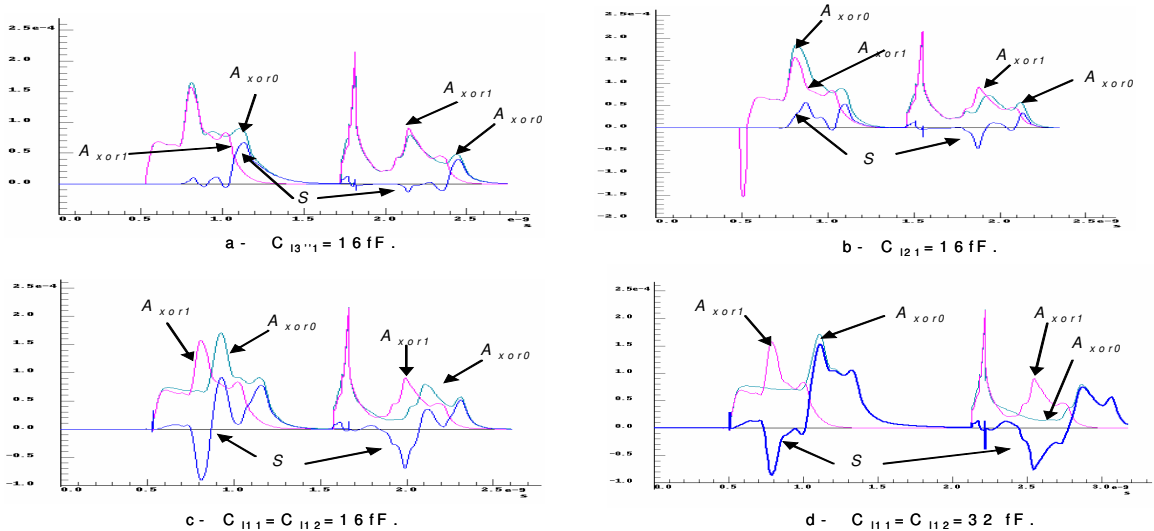

Fig. 7: Electrical Signature of the Dual-Rail Xor gate.
The net capacitances are varying from $C_d$=8fF to 32fF.

## VI. Improving place and route steps

We have demonstrated in paragraph V the effects of the net capacitances on the DPA bias signal. We now define a criterion for evaluating the net capacitance difference between the two rails of a dual-rail channel. If $C_{li}$ represents the net capacitance of rail $i$ ($i \in \{0,1\}$ for a dual rail channel) of channel $A$, then the dissymmetry of a channel "$A$" is defined by the following expression

$$d_A = \frac{|C_{l0} - C_{l1}|}{Min(C_{l0}, C_{l1})} \quad (12)$$

As illustrated in paragraph V, the lower the value of $d_A$, the more resistant to DPA the chip is. By using this criterion, we can then estimate the channels sensitivity to DPA.

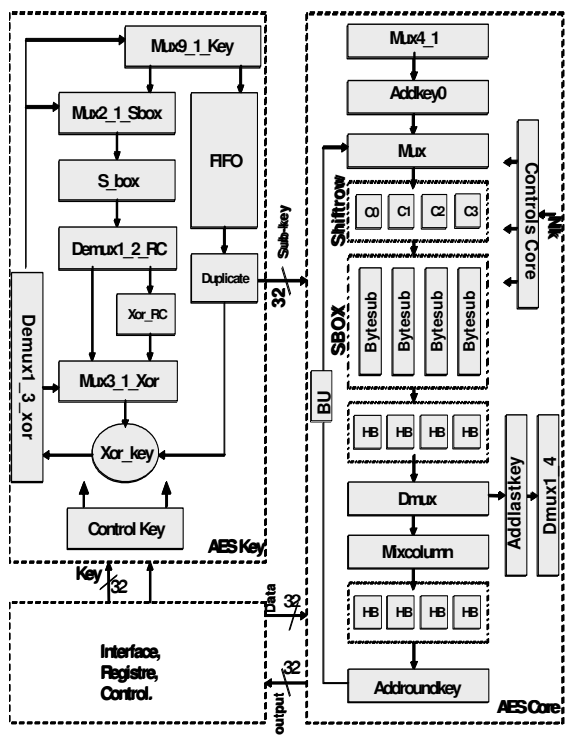

Fig. 8: Architecture of AES cipher block.

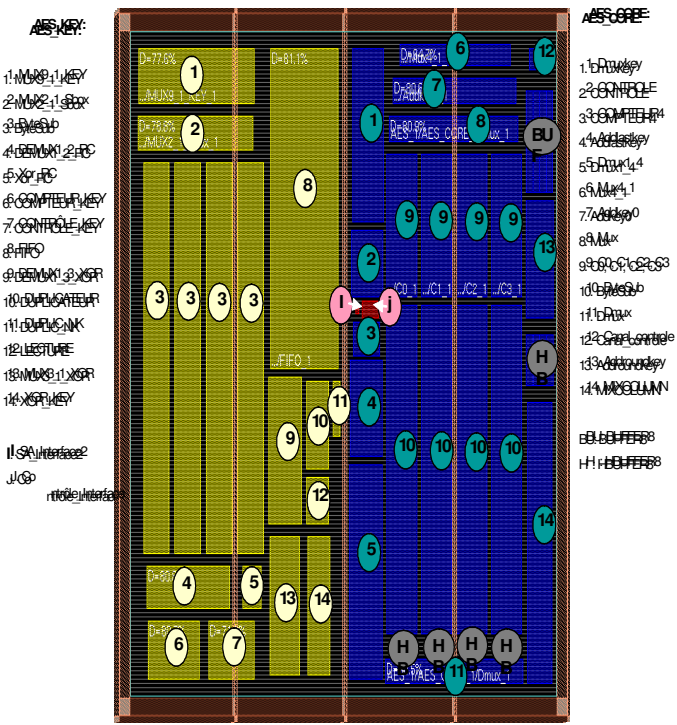

Fig. 9: Floorplan of the AES cipher block.
All blocks are constrained

Table 2: Most critical channels, i.e. presenting the highest value of the criterion, for versions 1 and 2.

| Version | AES_v1 - hierarchical | | | | | | | AES_v2 - flatten | | | | | | | |
|---|---|---|---|---|---|---|---|---|---|---|---|---|---|---|---|
| Block | Dmux of the AES core | | | | | | | HB block of the AES core | | | | | | | |
| Bit | 6 | | 21 | | 23 | | 25 | 25 | | 26 | | 30 | | 32 | |
| Dual-rail S0 \| S1 | 103 | 110 | 83 | 74 | 75 | 80 | 46 | 52 | 23 | 46 | 43 | 23 | 42 | 22 | 45 | 20 |
| $d_A$ | 0,07 | | 0,1 | | 0,06 | | 0,13 | 1 | | 0,86 | | 0,91 | | 1,25 | |

Traditionally, when the design is not too complex, the tools operate on a flat netlist which is more efficient than a hierarchical place and route in terms of area and performance. Unfortunately, the tool performs multiple random runs to optimize the design, in which the designer has no control on the net capacitances.

Therefore, we defined a place and route methodology which enables the designer to control the net capacitances and thus the criterion optimization.

The proposed approach is based on a hierarchical place and route flow which consists in dividing the design into small blocks and constraining their relative placement. The cells that implement a given function are gathered in a specified physical area which limits net length and dispersion.

This methodology is evaluated on the design of a secured asynchronous AES crypto-processor implementing the four-phase protocol, 1-of-N encoded data and balanced data paths [9]. The architecture of the AES crypto-processor is described in fig. 8. It is basically an iterative structure, based on three self-timed loops synchronized through communicating channels. Channel Sub-key synchronises the ciphering data-path with the sub-key computation data-path. The controller (finite state machine) generates signals which control both data-paths so that they execute Nr iterations as specified in the Rijndael algorithm [10]. Both ciphering data-path and Sub-key data-path are 32-bit wide. The details of the architecture choices are discussed in [9].

Fig. 9 represents the constrained floorplan of the AES chip (fig. 8) realized with Soc Encounter. To be able to quantify the benefit brought by the hierarchical place and route methodology, a flat AES Netlist is also routed to be used as a reference. Let us name AES_v1, the circuit designed with the proposed methodology, and AES_v2 the reference one (flat place and route).

The hierarchical approach has of course a cost in terms of silicon area. The core area of the first version (AES_v1 - hierarchical) is about 20% larger than the second version (AES_v2 - flatten).

To evaluate the benefits in terms of the DPA sensitivity, we computed for both versions the dissymmetry criterion for all the channels. Table 2 reports the most critical bits, i.e. the ones presenting the highest dissymmetry. For the second version (AES-v2 - flatten), the criterion value can reach up to 1.25. It represents an important source of leakage as demonstrated in equation 12 and illustrated in fig.7. Note that, even though most of the channels present a low criterion value, the existence of some channels having a high criterion value greatly degrades the DPA resistance level of the circuit. Moreover, we observed than the most sensitive channels are never the same from one place and route to another, confirming that the place and route process is not under the designer's control.

On the contrary, the hierarchical version (AES_v1) does not present any channel having a criterion value higher than 0.13 (Table 2). The channel net capacitance differences are therefore drastically reduced with the proposed approach.

## VII. Conclusion

This paper presented a logical and electrical formal analysis of the Differential Power Attack of secured Quasi Delay Insensitive asynchronous circuits. The definition of a formal model of the current dissipated in such circuits allowed us to apply DPA on this model, and thus identify sources of leakage. From this formal analysis, a design methodology is derived to limit the source of leakage by using hierarchical place and route. The gain obtained is illustrated with the design of a secure AES crypto-processor. It is shown that a significant improvement is achieved in theory, with the formal model.

The AES chip was sent for fabrication in August and will be back in October. DPA will be performed on the chip to characterize its resistance and verify the powerfulness of the method. The results will be presented at the conference.